\documentclass[pre,twocolumn,showpacs,superscriptaddress]{revtex4}

\usepackage{graphicx}
\begin{document}

\title {Sources and holes in a one-dimensional traveling wave
convection experiment}

\author {Luc Pastur}

\affiliation{Physics Department, Eindhoven University of Technology,
Postbus 513, 5600 MB Eindhoven, Netherlands}

\author{Mark-Tiele Westra}

\affiliation{Physics Department, Eindhoven University of Technology,
Postbus 513, 5600 MB Eindhoven, Netherlands}

\author{Daniel Snouck}

\affiliation{Physics Department, Eindhoven University of Technology,
Postbus 513, 5600 MB Eindhoven, Netherlands}

\author{Willem van de Water}

\affiliation{Physics Department, Eindhoven University of Technology,
Postbus 513, 5600 MB Eindhoven, Netherlands}

\author{Martin van Hecke}

\affiliation{Kamerlingh Onnes Laboratory, Universiteit Leiden,
Postbus 9504, 2300 RA Leiden, Netherlands}

\author{C. Storm}

\affiliation{Instituut-Lorentz, Universiteit Leiden, Postbus 9506,
2300 RA Leiden, Netherlands}

\author{Wim van Saarloos}

\affiliation{Instituut-Lorentz, Universiteit Leiden,
Postbus 9506, 2300 RA Leiden, Netherlands}

\date{\today}

\begin{abstract}
We study dynamical behavior of local structures, such as sources and
holes, in traveling wave patterns in a very long (2~m) heated wire
convection experiment. The {\em sources} undergo a transition from
stable coherent behavior to erratic behavior when the driving
parameter $\varepsilon$ is {\em decreased}. This transition, as well
as the scaling of the average source width in the erratic regime, are
both qualitatively and quantitatively in accord with earlier
theoretical predictions. We also present new results for the {\em
holes} sent out by the erratic sources.
\end{abstract}
~\vspace{-0.1mm}


\maketitle

\newcommand{\willem}[1]{{\bf willem} #1 \ {\bf willem}}

Traveling wave systems play an exceptional role within the field of
pattern formation.  If the transition to patterns is supercritical
(forward), the dynamics close to threshold should be amenable to a
description by the Complex Ginzburg-Landau (CGL) amplitude equation
\cite{CroHoh92}.  Theory and experiments are difficult to compare,
however, for the following two reasons: {\em{(i)}} both the CGL model
and experimentally observed traveling wave patterns exhibit an
astonishing variety of ordered, disordered and chaotic dynamics, which
can be difficult to characterize or compare. {\em{(ii)}} The dynamics
depends, in general, strongly on non-universal coefficients
\cite{chate,homoclons,hecke}, but the values of these coefficients are
difficult to determine in experiments \cite{croquette,liu,daviaud}.

The study of {\em local structures}, such as sources, fronts and holes
which play an important role in traveling wave systems
\cite{CroHoh92,chate,homoclons,hecke,daviaud,bot,kwaak,chif,SaaHoh92,vd},
provide a promising route to compare theory and experiment as they
partially circumvent these difficulties: their nontrivial behavior
often depends only on a {\em subset} of the coefficients
\cite{SaaHoh92} and is, in addition, relatively easy to characterize
experimentally \cite{daviaud,bot,kwaak,vd}.

\begin{figure}[t]
\includegraphics[width=8.6cm]{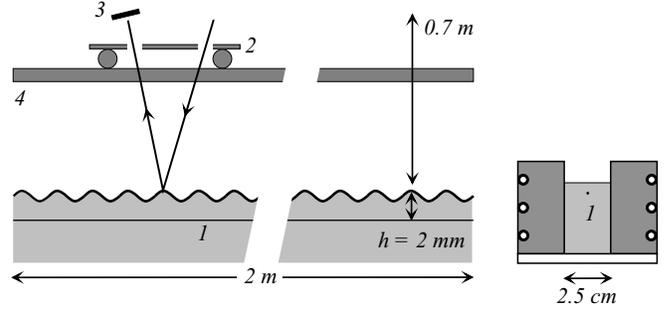}
\caption{Schematic side-view and cross section of the heated wire
experiment.  A thin (0.2~mm diameter) wire (1) is stretched beneath
the free surface of a fluid (depth $h = 2~$mm).  When it is heated by
sending an electrical current through it, surface waves are excited.
The slope of the waves is measured by reflecting a laser beam off the
surface onto a position sensitive detector (PSD) (3).  The laser and
the PSD are mounted on a cart (2) that rides on precision steel rods
(4). }\label{fig.setup}
\end{figure}

In this paper we present a successful example of this approach in a
heated wire convection experiment (Fig.~\ref{fig.setup}). This system
forms left and right traveling waves that suppress each-other; typical
states consist of patches of left and right traveling waves, separated
by sources (which send out waves) and sinks (which have two incoming
waves) \cite{snote}. Earlier theoretical work \cite{coullet}, which
was based on the amplitude equations (\ref{CCGLE1}-\ref{CCGLE2})
below, predicts that, essentially due to the transition from an
absolute to a convective instability \cite{absconv}, sources tend to
display a diverging width when the driving parameter $\varepsilon$ is
lowered beyond a critical value [Eq.~(\ref{epscrit})]. More recently
it was predicted that just before these stationary sources would
diverge, they become unstable and give way to {\em fluctuating
sources} of finite average width which display highly non-trivial
dynamics \cite{hecke}.

We indeed observe this nontrivial change in source behavior
when the driving (heating of wire) is decreased; not only the
measured transition value, but also the qualitative behavior
of sources is in accord with the predictions
\cite{coullet,hecke}. All properties necessary to compare
theory and experiment are measured in a set of independent
experiments. The fluctuating sources send out {\em holes}, and
we show that these display behavior very similar to that
predicted for homoclinic holes \cite{homoclons}.


\section{Experimental setup} Our experiment consists of a 2~m long
heated wire of diameter of $0.2\ {\rm mm}$ and resistivity of $50\
\Omega \ {\rm m}^{-1}$ that is placed under the free surface of the
fluid at a depth $h\! =\! 2\;{\rm mm}$ (see Fig.~\ref{fig.setup}). The
wire is stretched to the breaking limit and its maximum sagging is
$0.1\ {\rm mm}$.  The heat $Q$ dissipated in the wire drives the
system; through a combination of gravity- and surface tension induced
convection, surface waves emerge at $Q\! =\! Q_c$ that travel along the
wire \cite{vd}. The sides of our cell are made of brass and
contain copper tubes through which cooling water of $21.0\pm 0.1^\circ
C$ is circulated.  In order to guarantee a clean, uncontaminated free
surface, we use a low-viscosity, low-surface-tension silicon oil
\cite{oil}.

A sensitive linear measurement of the surface slope along the cell is
obtained by recording the reflection of a laser beam off the fluid
surface onto a position sensitive device. Both laser and position
detector are mounted on a computer-controlled cart which travels on
precision stainless steel rods. This allows us to measure surface wave
amplitudes as small as $0.5\;\mu{\rm m}$.  The signals of the scanning
device are 
wave frequency and Hilbert transformed to yield the complex valued
field $A(x,t)=|A|\exp(i \phi)$.  From this the local wave number is
computed as $q(x,t) = \partial \phi(x,t) / \partial x$.
%
%
To improve the signal to noise ratio, running averages over a time
interval of 10~s are performed.

Vince and Dubois \cite{vd} already demonstrated that the primary
bifurcation at $Q = Q_c$ is supercritical and explored the phase
diagram as a function of $Q$ and wire depth $h$.
For $\varepsilon \lesssim 0.15$ the amplitude exhibits the
scaling $|A| \sim \varepsilon^{1/2}$.  This is expected near a
supercritical bifurcation, and it also sets the range of
applicability of the amplitude description.


\section{Amplitude Equations}
For systems with counter-propagating waves, the appropriate amplitude
equations are the coupled one-dimensional CGL equations \cite{CroHoh92}:
\newline
\begin{eqnarray}
\label{CCGLE1}
   \tau_0\!\!\! &( \partial_t A_{R} \!+\! s_0\partial_x A_{R}\ ) \!=\!
   \varepsilon A_{R} \!+\!\xi_0^2 (1 \!+\! i c_1) \partial_x^2 A_{R}
   \nonumber\\ &\!-\! g_0 (1 \!-\! i c_3) |A_{R}|^2 A_{R} \!-\! g_2 (1 \!-\!
   i c_2) |A_{R}|^2 A_{L},\\ \tau_0 \!\!\! &( \partial_t A_{L} \!-\!
   s_0\partial_x A_{L}\ ) \!=\! \varepsilon A_{L} \!+\!\xi_0^2 (1 \!+\! i
   c_1) \partial_x^2 A_{L} \nonumber\\ &\!-\! g_0 (1 \!-\! i c_3) |A_{L}|^2
   A_{L} \!-\! g_2 (1 \!-\! i c_2) |A_{L}|^2 A_{R}~.\label{CCGLE2}
\end{eqnarray}
Here $A_R$ and $A_L$ are the amplitude of the right and left moving
waves, $s_0$ is the linear group velocity, and $c_1,c_2$ and $c_3$
measure the linear and nonlinear dispersion. The experimentally
accessible control parameter $\varepsilon$ measures the distance from
threshold.  The coefficients $\tau_0,\xi_0$ and $g_0$ give the scales
of time, space and amplitude.  To model our experiment, where left and
right traveling waves suppress each-other, $g_2$ should be larger than
$g_0$ \cite{hecke}.

\begin{figure}[t]
\includegraphics[width=8.6cm]{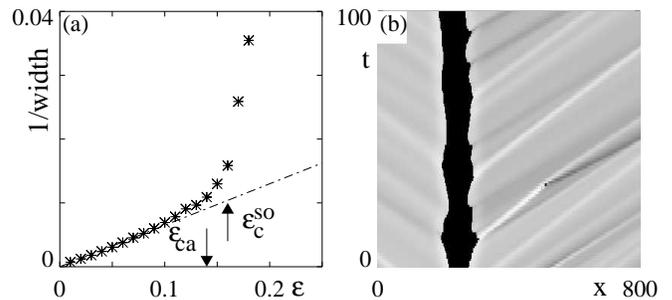}
\caption{ Numerical results for the behavior of sources in the coupled
amplitude equation (\ref{CCGLE1}), (\ref{CCGLE2}). (a) Inverse average
source width as a function of $\varepsilon$, for the coupled CGL
equations with $s_0\!=\!1.5,c_1\!=\!-1.7,c_2\!=\!0,c_3\!=\!0.5,
g_0\!=\!1$ and $g_2=2$. The coefficients $s_0$ and $c_1$ were chosen
to be similar to those measured in the experiment; also $g_2>g_0$ in
the experiment. The values of $c_2$ and $c_3$ where chosen such that
the plane waves remain stable; their precise value does not play a
significant role then.  Note the crossover near
$\varepsilon_{ca}=0.14$. (b) Space-time plot of the local wave number
of a fluctuating source for
$\varepsilon\!=\!0.11<\varepsilon_c^{so}\approx 0.14$, illustrating
fluctuations of the width. In the black region the amplitude has
fallen below 10\% of the saturated value; the light and dark curves
correspond to hole-like wave number packets send out by the source.}
\label{fromphysd}
\end{figure}

\subsection{Scaling}
Sources show complicated behavior within the amplitude equations
(\ref{CCGLE1}-\ref{CCGLE2}) \cite{coullet,hecke}.  For
\begin{equation}
\varepsilon > \varepsilon_c^{so} \gtrsim \varepsilon_{ca} = \frac{
(s_0 \tau_0 / \xi_0)^2}{4 (1+c_1^2)} \label{epscrit}
\end{equation}
sources are coherent structures with a well-defined shape, while for
$\varepsilon < \varepsilon_c^{so}$ sources start to fluctuate and
their average width scales as $\propto \varepsilon^{-1}$ (see
Fig.~\ref{fromphysd}). The quantity $\varepsilon_{ca}$ in
Eq. (\ref{epscrit}) is simply the value of $\varepsilon$ where the
transition from absolute to convective instability of the $A \equiv 0$
state occurs \cite{coullet,absconv}; its relevance can be understood
as follows. Consider the dynamics of a single front in the left-moving
wave amplitude $A_L$ only, for which $A_L(x\gg 1) \rightarrow 0$. The
propagation velocity of this front is given by a competition between
the linear group-velocity, which tends to convect any structure to the
{\em left} with velocity $s_0$, and the propagation of the front into
the $A=0$ state with, in the comoving frame, velocity
$v^*\!=\!2 \xi_0 \sqrt{\varepsilon(1+c_1^2)}/ \tau_0$
\cite{hecke,coullet}; the front velocity in the lab frame is thus
$v^*-s_0$. Viewing a source as a pair of fronts in $A_L$ (on the left)
and $A_R$ (on the right), it is clear that these fronts move together
when $\varepsilon > \varepsilon_{ca}$, but move apart when
$\varepsilon < \varepsilon_{ca}$; the change in direction of front
propagation precisely corresponds in the transition from absolute to
convective instability.

\begin{figure}
\includegraphics[width=8.6cm]{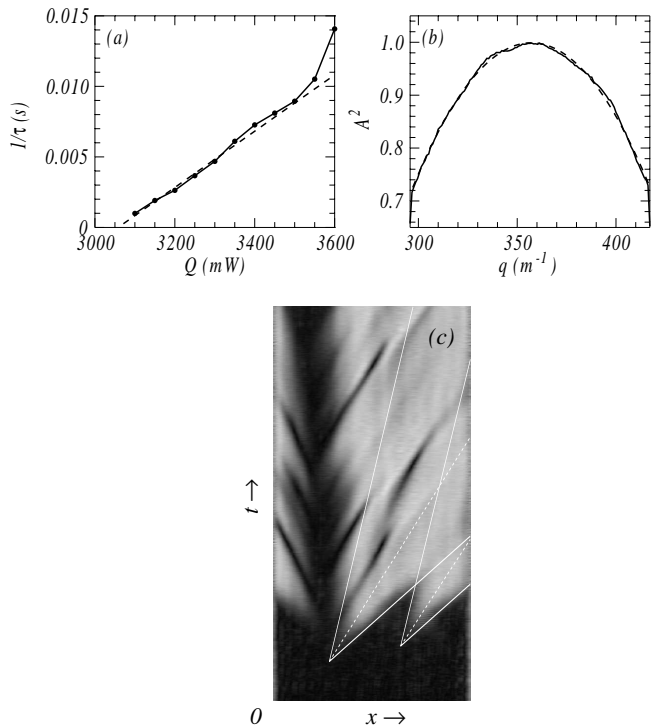}
\caption{Determination of the coefficients of the CGLE.  (a) Timescale
$\tau$ determined from the exponential growth of the amplitude in
quench experiments for various value of the heating power $Q$.  This
time scales as $\tau = \tau_0 / \varepsilon$, with $\tau_0 = $16(1)~s.
(b) Correlation length $\xi_0$. Full line: histogram of squared
modulus $|A|^2$ vs $q$ which is measured from a modulated wave field
at $\varepsilon = 0.10$.  Dashed line: fit of $|A|^2 = 1 -
\xi_0^2/\varepsilon (q - q_0)^2$, with $\xi_0 = 2.7(6)\times
10^{-3}\;{\rm m}$. $|A|$ is normalized so that $|A|=1$ corresponds to
waves with wave number $q_0$.  (c) Front velocity.  Shown is the
modulus $|A(x,t)|$.  The $x-$extent of the scan is 0.682~m, the total
time is 5242~s. At $t = 0$ the power is quenched from $Q = 0$ to $Q =
(1 + \varepsilon) Q_c$, with $\varepsilon = 0.051$.  The white lines
outline two fronts. Since $\varepsilon < \varepsilon_c^{so}$ here,
both fronts propagate in the same direction (here to the right).}
\label{fig.coef}
\end{figure}

Numerical simulations of Eq.\ \ref{CCGLE1} were done in order to see
whether the experimentally observed source behavior described below
could be understood on basis of the amplitude description.  Such
simulations \cite{hecke} have revealed that sources do {\em not}
simply move apart and diverge when the instability of the $A \equiv 0$
state becomes convective; for $\varepsilon=\varepsilon_c^{so}\gtrsim
\varepsilon_{ca}$, when the sources have become very wide, they start
to fluctuate.  For smaller $\varepsilon$, the average source width
scales as $\varepsilon ^{-1}$ (see Fig.~2). The mechanism responsible
for the sources staying at a finite but large average width is not
completely understood and may depend on the noise strength. In the low
noise limit, the ``tip'' regions of the two fronts sense the other
mode which leads to the formation of phaseslips there. The resulting
perturbations are then advected by the group-velocity and amplified by
the linear growth-rate, resulting in a jittery motion of the
front. For larger noise strength, convective amplification of noise
may compete with this mechanism.

These phenomena are illustrated in Fig.\ \ref{fromphysd} which
has been calculated for parameter values which are in the
range of the experimental ones, but we emphasize that the
predicted source instability is generic and insensitive to the
precise parameter values.

\section{Measurements} 

\subsection{Front and group velocities} 

Now that we have discussed the theoretical predictions for sources, we
return to our experiment. For a comparison of the source behavior with
theory, we need to determine the transition from convective to
absolute instability, which requires measurements of the group
velocity and the front velocity as function of $\varepsilon$.

The group velocity $s_0$ was determined from the propagation of
deliberate perturbations of the surface.  We found that it has the
same sign as the phase velocity and that is shows only a weak
$q-$dependence, so we associate the measured value $ 2.1(1) \times
10^{-4} \; {\rm m \: s}^{-1}$ with the linear group velocity $s_0$.

Fronts where made by quenching the heating power $Q$ to a
finite value $Q= (1+ \varepsilon) Q_c$ at $t\!=\!0$. After a
short while, waves invade the unstable surface in the form of
fronts. The boundaries of these fronts travel with $s_0 \pm
v_f$ respectively, where $v_f$ scales with $\varepsilon$ as
$v_{f0} \sqrt{\varepsilon}$. Fig.\ \ref{fig.coef}c. shows the
evolution of the amplitude of the waves for $\varepsilon
=0.051$; this value appears to be below $\varepsilon_{ca}$
because the velocity of the fronts has the same sign as the
group velocity. The results of several experiments, both at
$\varepsilon < \varepsilon_{ca}$ and $\varepsilon >
\varepsilon_{ca}$ yields that $v_{f0}=5.4(5) \times
10^{-4}$ m/s. Comparing this to the value obtained for $s_0$,
we immediately find that $\varepsilon_{ca} \approx 0.15(5)$. A
alternative estimate of $\varepsilon_{ca}$ was made from
observing at which $\varepsilon$ the slowest moving edge of a
front has zero velocity, which led to a value of
$\varepsilon_{ca} = 0.10(2)$. 

\subsection{Measurements of the coefficients}
In principle, a confrontation of theory and experiment can also be
performed by measuring the characteristic time ($\tau_0$) and length
($\xi_0$) scales, the linear group velocity $s_0$ and linear
dispersion coefficient $c_1$, since from these the transition from
convective to absolute instabilities also follows (see Eq.~3). Note
that starting from the full hydrothermal equations, these coefficients
can in principle be obtained from a systematic amplitude expansion
\cite{CroHoh92}.  At present, we can only obtain $c_1$ via
measurements of the front-velocity which leads to a consistency check
(see below). The length and time scales are relevant for comparing
spacetime diagrams of experiment and theory and are measured
independently.


The characteristic time is determined from measurements in which the
growth of the amplitude is followed when, after a sufficient long
transient in which a plane wave is established, $\varepsilon$ is
increased from $\varepsilon=0.017$ to larger values. The initial
growth of $|A|$ is exponential $\sim \exp(t/\tau)$, and repeating this
experiment for various values of $\varepsilon$ yields the data
presented in Fig.~\ref{fig.coef}. Using that $\tau$ scales as $\tau =
\tau_0 / \varepsilon$, we obtain that $\tau_0 = $16(1)~s.

The length scale $\xi_0$ was measured from weakly modulated waves in
the single-wave domains; according to Eq.\ (\ref{CCGLE1}) with
$A_L=0$, these are related by $ A(q)^2 g_0 = \left( \varepsilon -
\xi_0^2 q^2 \right)$. Plotting these values we obtain
Fig.~\ref{fig.coef}(b), in which we recognize the quadratic behavior
of $|A|$ as a function of $q$.  The measured $\xi_0$ differed slightly
(but not systematically) from run to run and from a series of such
measurements and fits we find for the correlation length $\xi_0 = (2.7
\pm 0.6) \times 10^{-3} \;{\rm m}$, which only close to threshold
becomes similar to the basic wavelength of the traveling waves.

Taking these time and length scales and the measured front
velocity $v_{f0}$ used before, we find then that $( 1
\!+\!  c_1^2 ) \!\approx\! 2.6 $, from which it follows that
$c_1 = \pm 1.3(4)$. A weak consistency check is that
$\left( 1 + c_1^2 \right)$ should be larger than one;
independent measurements of $c_1$ would lead to a stronger
consistency check.


\begin{figure}
\includegraphics[width=8.6cm]{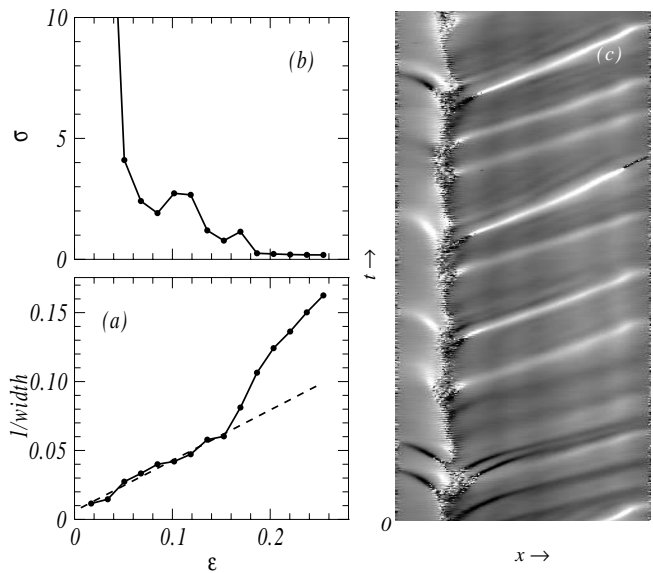}
\caption[]{(a) Dependence of the width of a source on the reduced
control parameter $\varepsilon$. Dots: mean width $\langle w
\rangle^{-1}$, dashed line: fit of $\langle w \rangle \sim
\varepsilon^{-1}$.
%
(b) Dependence of the rms fluctuation $\sigma=\sqrt{<w^2>-<w>^2}$ of
the width of a source on $\varepsilon$.  Notice that the source
becomes unstable for $\varepsilon > \varepsilon_{ca}$.  Note that in
(a)-(b), length scales have been non-dimensionalized by the
characteristic scale $\xi_0$.
(c) Space-time diagram of the wave number field $q(x,t)$ of an
unstable source, $\varepsilon = 0.11$; the extent of the $x-$axis is
$158\: \xi_0$, the total time is $660\: \tau_0$.}
\label{spacetimeplots}
\label{fig.width}
\end{figure}

\section{Comparison between experiment and theory}
\subsection{Sources} 
Now that all relevant parameters of the amplitude equations are
approximately known, we turn to the behavior of sources in our
experiment. The dependence of the width $w(t)$ of sources on the
control parameter $\varepsilon$ was measured in long experimental runs
in which a source was located at large heating power $\varepsilon
\approx 0.3$ and then followed at progressively smaller values of
$\varepsilon$ \cite{widthnote}.  At each $\varepsilon$, the source was
observed for several hours by scanning the fluid surface, while
keeping the experimental conditions constant.

From the width $w(t_i)$ at discrete scan times $t_i$ we computed the
mean $\langle w \rangle$ (as well as the standard deviation $\sigma$
\cite{widthnote}).  Fig.\ \ref{fig.width} shows that the behavior of
$\langle w \rangle$ as a function of $\varepsilon$ in the experiments
shows the same qualitative features as the numerical simulations of
Fig. \ref{fromphysd}: For decreasing $\varepsilon$ the width appears
to diverge, but at $\varepsilon \approx 0.15$ there is a 
%
cross-over to a $\langle w \rangle \sim \varepsilon^{-1}$ behavior.
Below this crossover value, the sources fluctuate strongly and the
standard deviation of the width rapidly increases \cite{pastur}. In a
cyclic fashion, these sources first grow and spawn outward-spreading
wave fronts, leaving an interval of near-zero wave amplitude behind
in the source core. Here phase slips occur, which make the fronts
jump back; the resulting phase twists are carried away by hole
structures which travels roughly with the group velocity (the light
and dark lines).  In our numerical simulations of of the coupled CGL
equations (Fig.~\ref{fromphysd}) exactly the same hole structures are
observed.
The crossover {\em value} for $\varepsilon_c^{so}$ is consistent with
the transition value $\varepsilon_{ca}$ that we determined before.

\begin{figure}
\includegraphics[width=8.6cm]{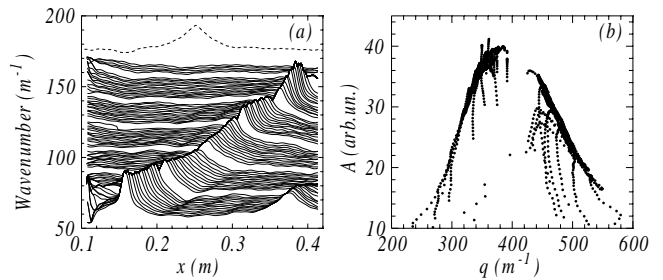}
\caption[] {(a) Wave number profile of a hole emitted from the unstable
source of Fig.\ \protect\ref{fig.width}b.
Dashed line: typical wave number profile.  To help reading wave number
off the vertical axis, the plot has been sectioned.
(b) Scatter-plot of the minimum of the modulus versus the extreme (in
$x$) of the wave number along each of the holes shown in Fig.
\protect\ref{fig.width}b. Both compression (large $q$) and dilation
(small $q$) holes belong to a one-parameter family.}
\label{clon}
\end{figure}

\subsection{Holes} The
structures sent out by the erratic sources display a dip in
the amplitude $|A|$ and are therefore referred to as holes. It
is well known that holes play an important role in the
dynamics of traveling wave systems, and that different types
can be distinguished by whether the wave numbers of their two
adjacent waves are similar or substantially different
\cite{bot,daviaud,chate,homoclons,SaaHoh92}. From the measured
wave number profile in Fig.~\ref{clon}a it can be seen that the
wave numbers at the back and front side of the hole are very
similar. We therefore associate these holes with so-called
{\em homoclinic holes} \cite{homoclons}.  In addition, they display
the following typical homoclinic hole behavior (see Fig.~2b and 4b):
they do not send out waves and occur quite close together,
they can evolve to defects and their propagation velocity
(which in lowest order is given by $s_0$) depends on the value
of the extremum of $q$. In the local wave number plot of Fig.\
\ref{fromphysd} dilation holes (the dark lines) have a larger
velocity than compression holes (light lines), just as in the
experimental plot Fig.\ \ref{spacetimeplots}c. In fact, the
correlation between the type of wave number modulation and the
velocity of these coherent structures depends on the sign of
$c_1$, which was selected accordingly for the numerical
simulations.

Since homoclinic holes are dynamically unstable, their local profiles slowly
evolve along a one-parameter family; on a scatter plot of the values
of the minimum of $|A|$ versus the corresponding extremum of $q$,
these values collapse on a single curve \cite{homoclons}. The holes in
our experiment precisely show this behavior: The extrema of $|A|$ and
$q$ rapidly evolve toward a parabolically shaped curve, and stay there
during their further evolution (Fig.~\ref{clon}b). We only observed
these holes in our experiment for at most a few characteristic times
--- too short to see clear signs of the weak instability predicted
from the CGL equation.

\section{Discussion and outlook}
Our experiments raise a number of suggestions for further work.
{\em{(i)}} The width where sources start to fluctuate is larger
in the theory ($O(100 \:\xi_0)$) than in experiments ($O(20
\:\xi_0)$), while the fluctuations appear stronger for experimental
sources. Experimental noise or non-adiabatic effects which perturb
the fronts may play a role here.  {\em{(ii)}} Earlier experiments
\cite{vd} have shown that for different heights of the wire,
qualitatively different behavior occurs.  Systematic measurements of
the coefficients as a function of height may turn the heated wire
experiment into a CGLE-machine with tunable coefficients.
{\em{(iii)}} Longer observations and more controlled generation of
holes may shed more light on their relation to the homoclinic holes
predicted by theory, and may show the highly characteristic
divergence of lifetime as a function of initial condition
\cite{homoclons}. {\em{(iv)}} Sinks show non-adiabatic phase-matching
and in fact posses completely anti-symmetric profiles \cite{pastur};
there is no clear theoretical understanding of this.

{\em Acknowledgment}  We thank Ad Holten and Gerald Oerlemans
for technical assistance. Financial support by EC Network
Contract FMRX-CT98-0175, by the ``Nederlandse Organisatie voor
Wetenschappelijk Onderzoek (NWO)'' and by ``Stichting
Fundamenteel Onderzoek der Materie (FOM)'' is gratefully
acknowledged.


\begin{thebibliography}{99}
\bibitem{oil}
The brand name of the oil is Tegiloxan 3, produced by Goldschmidt AG
(Essen, Germany). At $21^\circ C$ it has viscosity $\nu =3.397 \cdot
10^{-6}\ m\ s^{-2}$, density $\rho =892.4\ kg\ m^{-3}$, surface
tension $\alpha =18.3\cdot 10^{-3}\ N\ m^{-1}$, with temperature
coefficient ${\rm d}\alpha/{\rm d}T = -9.7 \cdot 10^{-5}\ N\ m^{-1}\
K^{-1}$, and a refractive index $n = 1.395$.
\bibitem{widthnote} To quantify the width of sources we use our
finding that the envelope of the experimental pattern is well fitted
by the expression $ S_+(x - x_0) + S_-(x - x_0)$, with $ S_\pm \propto
\left( 1 + a \exp(\pm 2 b x) \right)^{-1/2}$. The source width $w$ was
defined as $S_\pm(\mp w/2) = 1/2$.
\end{thebibliography}
\end{document}